\documentstyle[sprocl,epsfig,amssymb]{article}

\def\NPB{{\em Nucl.\ Phys.} B}
\def\PLB{{\em Phys.\ Lett.}  B}
\def\PRL{\em Phys.\ Rev. Lett.}
\def\PRD{{\em Phys.\ Rev.} D}
\def\EPJ{{\em Eur.\ Phys.\ J.} C }
\def\be{\begin{equation}}
\def\ee{\end{equation}}
\def\bea{\begin{eqnarray}}
\def\eea{\end{eqnarray}}
\begin{document}

\title{CP VIOLATION IN SUPERSYMMETRY\footnote{%
Work supported by the `Fonds zur F\"orderung der
wissenschaftlichen For\-schung' of Austria, FWF Project No.~P16592-N02
and by the European Community's Human Potential Programme
under contract HPRN-CT-2000-00149.}\\}

\author{ STEFAN HESSELBACH }

\address{Institut f\"ur Theoretische Physik, Universit\"at Wien, 
  A-1090 Vienna, Austria}


%%%%%%%%%%%%%%%%%%%%%%%%%%%%%%%%%%%%%%%%%%%%%%%%%%%%%%%%%%%%%%%%%%%%%%%%
% Cover Page                                                           %
%%%%%%%%%%%%%%%%%%%%%%%%%%%%%%%%%%%%%%%%%%%%%%%%%%%%%%%%%%%%%%%%%%%%%%%%

\thispagestyle{empty}
\setcounter{page}{0}
\renewcommand{\thefootnote}{\fnsymbol{footnote}}

{\hspace*{\fill} UWThPh-2004-27}

{\hspace*{\fill} LC-TH-2004-014}

{\hspace*{\fill} hep-ph/0409192}

\vfill
\begin{center}

{\Large\bf
CP VIOLATION IN SUPERSYMMETRY\footnote{%
Contribution to the proceedings of the International Conference on
Linear Colliders (LCWS~04), Paris, April 19 -- 23, 2004.}
}

\vspace{14mm}

{\large
STEFAN HESSELBACH
}

\vspace{3mm}

{\it Institut f\"ur Theoretische Physik, Universit\"at Wien, 
  A-1090 Vienna, Austria}

\end{center}

\vfill

\begin{abstract}
A review about some recent
studies on the determination of the CP-violat\-ing complex parameters
in supersymmetry at an $e^+ e^-$ linear collider is presented.
CP-even observables, like masses, cross sections and branching ratios,
can have a strong dependence on the supersymmetric phases. However,
CP-odd observables, like asymmetries based on triple product
correlations, 
are necessary to unambiguously establish CP violation.
In the chargino and neutralino sector these asymmetries can be as
large as 30\,\%  and will therefore be an important tool for the
search for CP-violating effects in supersymmetry.
\end{abstract}

\vfill

\newpage

\setcounter{footnote}{0}
\renewcommand{\thefootnote}{\alph{footnote}}

%%%%%%%%%%%%%%%%%%%%%%%%%%%%%%%%%%%%%%%%%%%%%%%%%%%%%%%%%%%%%%%%%%%%%%%%

\maketitle\abstracts{
A review about some recent
studies on the determination of the CP-violating complex parameters
in supersymmetry at an $e^+ e^-$ linear collider is presented.
CP-even observables, like masses, cross sections and branching ratios,
can have a strong dependence on the supersymmetric phases. However,
CP-odd observables, like asymmetries based on triple product
correlations, 
are necessary to unambiguously establish CP violation.
In the chargino and neutralino sector these asymmetries can be as
large as 30\,\%  and will therefore be an important tool for the
search for CP-violating effects in supersymmetry.
}

%***********************************************************************
\section{Introduction}
%***********************************************************************

The small amount of CP violation in the Standard Model (SM), which is
caused by the phase in the Cabibbo-Kobayashi-Maskawa matrix, is not
sufficient to explain the baryon-antibaryon asymmetry of the
universe \cite{bau}.
The Lagrangian of the Minimal Supersymmetric Standard Model (MSSM)
contains several complex parameters, which can
give rise to new CP-violating phenomena \cite{mssmcpv}.
After eliminating unphysical phases two complex parameters remain in
the neutralino and chargino sector, the U(1) gaugino mass parameter
$M_1$ and the higgsino mass parameter $\mu$.
Furthermore the SU(3) gaugino mass parameter $M_3$ and the trilinear
scalar couplings $A_f$ in the sfermion sector can be complex.

The phases of the complex parameters are constrained or
correlated by the experimental upper limits on the electric
dipole moments of electron, neutron and the atoms
${}^{199}$Hg and ${}^{205}$Tl \cite{edmexp}.
For example, the phase of $\mu$ is restricted to 
$|\phi_\mu| \lesssim 0.1\pi$ in mSUGRA-type models.
However, there
may be cancellations between the contributions of different
complex parameters, which allow larger values for the phases
\cite{edm,Choi:2004rf}.
Moreover, the restrictions are very model dependent.
For example, when also lepton flavor violating terms
are included, then the restriction on $\phi_\mu$
may disappear \cite{Bartl:2003ju}.

In this contribution I will summarize several projects how to
determine the supersymmetric (SUSY) phases with help of CP-even and CP-odd
observables at a future $e^+ e^-$ linear collider.

%***********************************************************************
\section{CP-even Observables}
%***********************************************************************

\begin{figure}[t]
\setlength{\unitlength}{1cm}
\begin{picture}(11.9,4.6)
%\put(0,0){\framebox(11.9,4.6){}}

\put(0,4.2){(a)}
\put(0.1,0){\epsfig{file=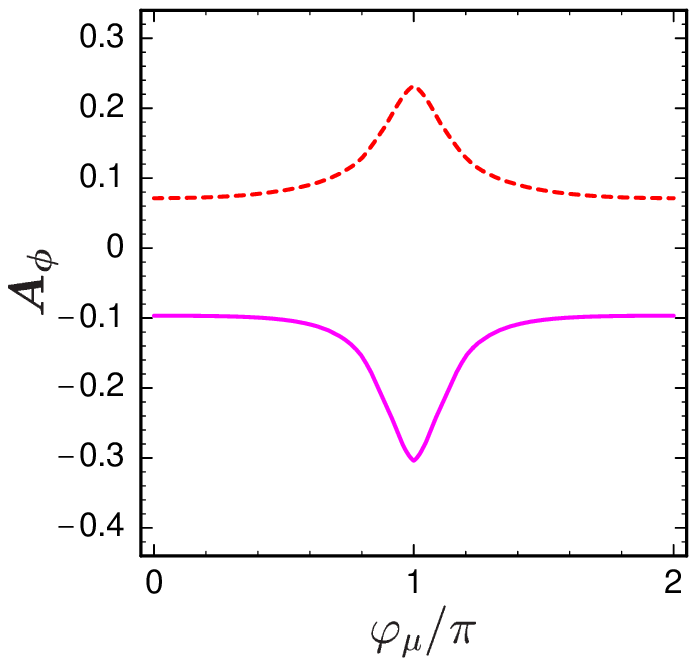, width=5cm}}

\put(5.5,4.2){(b)}
\put(6,0.05){\epsfig{file=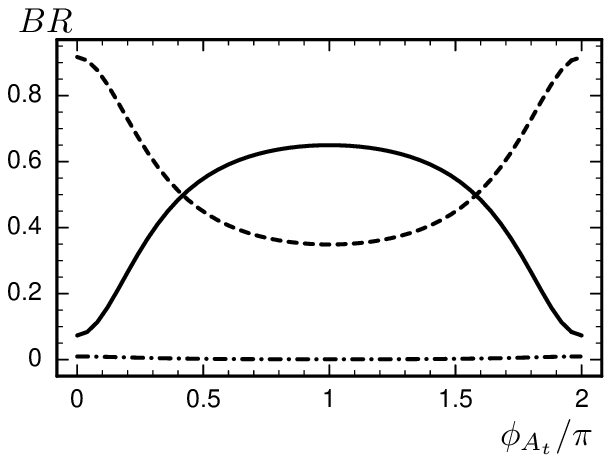, scale=0.96}}

\end{picture}

\caption{\label{fig:CPeven}
(a) Azimuthal asymmetry $A_\phi$ for $e^+ e^- \to \tilde{\chi}^+_1
\tilde{\chi}^-_2$ with the subsequent decays 
$\tilde{\chi}^-_j \to \ell^- \tilde{\nu}_\ell$ (solid) and
$\tilde{\chi}^-_j \to W^- \tilde{\chi}^0_1$ (dashed) for
$\sqrt{s} = 800$~GeV, transverse beam polarizations $P^\perp_{e^-} = 0.8$,
$P^\perp_{e^+} = 0.6$ and 
$|\mu| = 400$~GeV, $M_2 = 200$~GeV,
$|M_1|/M_2 = 5/3 \, \tan^2\theta_W$,
$\phi_{M_1} = 0$,
$\tan\beta = 3$, $m_{\tilde{\nu}} = 150$~GeV.
From \protect\cite{Bartl:2004xy}.
(b) Branching ratios $BR$ of the decays
$\tilde{t}_1 \to \tilde{\chi}^+_1 b$ (solid),
$\tilde{t}_1 \to \tilde{\chi}^0_1 t$ (dashed) and
$\tilde{t}_1 \to W^+ \tilde{b}_1$ (dashdotted)
for $\tan\beta = 6$, $M_2=300$~GeV,
%$|M_1|/M_2 = 5/3 \, \tan^2\theta_W$,
$|\mu|=350$~GeV,
$|A_b|=|A_t|=800$~GeV,
$\varphi_\mu=\pi$, $\varphi_{M_1}=\varphi_{A_b}=0$,
$m_{\tilde{t}_1}=350$~GeV, $m_{\tilde{t}_2}=700$~GeV,
$m_{\tilde{b}_1}=170$~GeV, $M_{\tilde{Q}}>M_{\tilde{U}}$
and $m_{H^\pm}=900$~GeV.
From \protect\cite{squarkpapers}.}
\end{figure}

The study of production and decay of 
charginos ($\tilde{\chi}^\pm_i$) and neutralinos ($\tilde{\chi}^0_i$)
and a precise determination of the underlying SUSY
parameters $M_1$, $M_2$, $\mu$ and $\tan\beta$
including the phases $\phi_{M_1}$ and $\phi_\mu$
will play an important role at future linear colliders
\cite{LC}.
In \cite{NeuChaParDet} methods to determine these parameters based on
neutralino and chargino mass and cross section measurements have been
presented.
In \cite{Choi:2004rf} the impact of the SUSY phases on chargino,
neutralino and selectron production has been analyzed and
significances for the existence of non-vanishing phases have been defined. 
In \cite{Bartl:2004xy} chargino production 
$e^+ e^- \to \tilde{\chi}^+_i \tilde{\chi}^-_j$
at a linear collider with transversely polarized beams has been
analyzed. It is shown that CP-odd triple product correlations
involving the transverse beam polarizations vanish, if at least one
subsequent chargino decay is not observed. However, 
for subsequent chargino decays
$\tilde{\chi}^-_j \to \ell^- \tilde{\nu}_\ell$ or
$\tilde{\chi}^-_j \to W^- \tilde{\chi}^0_1$
it is possible to define CP-even azimuthal asymmetries,
which have a strong dependence on $\phi_\mu$ and $\phi_{M_1}$
(Fig.~\ref{fig:CPeven}~(a)).
Further methods to probe the CP properties of neutralinos 
are described in \cite{Choi:2004jf}.

In contrast to the parameters of the chargino and neutralino sector
it is more difficult to measure the trilinear couplings $A_f$ in the
sfermion sector.
Cross section measurements of sfermion production processes allow the
determination of the sfermion masses and mixing angles
which in turn allow the determination of the
parameters $A_f$ in the real case \cite{Bartl:2000kw}.
In \cite{staupapers,squarkpapers,Gajdosik:2004ed} the impact of the CP
phases of $A_\tau$, $A_t$, $A_b$, $\mu$
and $M_1$ on production and decay of $\tilde{\tau}_{1,2}$,
$\tilde{\nu}_\tau$, $\tilde{t}_{1,2}$ and $\tilde{b}_{1,2}$
have been studied.
The branching ratios of fermionic decays of $\tilde{\tau}_{1}$ and
$\tilde{\nu}_\tau$ show a significant phase dependence for $\tan\beta
\lesssim 10$ whereas it becomes less pronounced for $\tan\beta > 10$.
The branching ratios of $\tilde{\tau}_{2}$ into Higgs bosons
depend very sensitively on the phases for $\tan\beta \gtrsim 10$.
The branching ratios of $\tilde{t}_{1,2}$ show a pronounced phase
dependence in a large region of the MSSM parameter space
(Fig.~\ref{fig:CPeven}~(b)).
In the case of $\tilde{b}_{1,2}$ decays there can be an appreciable
$\varphi_{A_b}$ dependence, if $\tan\beta$ is large and the decays
into Higgs bosons are allowed.
Further the expected accuracy in determining the SUSY
parameters has been estimated by a global fit of measured masses, branching
ratios and production cross sections. $A_\tau$, $A_t$ and $A_b$
can be expected to be measured with 10\,\%, 2 -- 3\,\% and 50\,\%
accuracy, respectively,
$\tan\beta$ with 1\,\% (2\,\%) accuracy in case
of small (large) $\tan\beta$ and the other parameters with
approximately 1\,\% accuracy.

%***********************************************************************
\section{CP-odd Observables}
%***********************************************************************

In order to unambiguously establish CP violation in SUSY,
including the signs of the phases, the use of CP-odd
observables is inevitable.
In SUSY T-odd triple product correlations between momenta and spins of the
involved particles allow the definition of CP-odd asymmetries already
at tree level \cite{Choi:1999cc}.
Such asymmetries have been analyzed for neutralino and chargino
production with subsequent three-body \cite{Bartl:2004jj}
and two-body decays \cite{AT2body,Bartl:2004vi}, where full
spin correlations between production and decay have to be included
\cite{spincorr}.

A Monte Carlo study of T-odd asymmetries in selectron and neutralino
production and decay including initial state radiation,
beamstrahlung, SM backgrounds and detector effects
has been given in \cite{MonteCarlo}.
It has been found that asymmetries $A_T \sim 10\,\%$ are detectable
after few years of running of a linear collider.

\begin{figure}[t]
\setlength{\unitlength}{1cm}
\begin{picture}(11.9,5)
%\put(0,0){\framebox(11.9,5){}}

\put(0,4.6){(a)}
\put(0.3,0){\epsfig{file=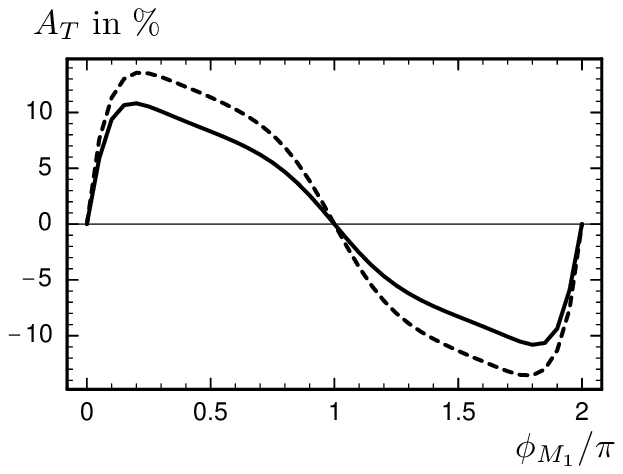,scale=1.03}}

\put(7.1,4.6){(b)}
\put(7.6,0){\epsfig{file=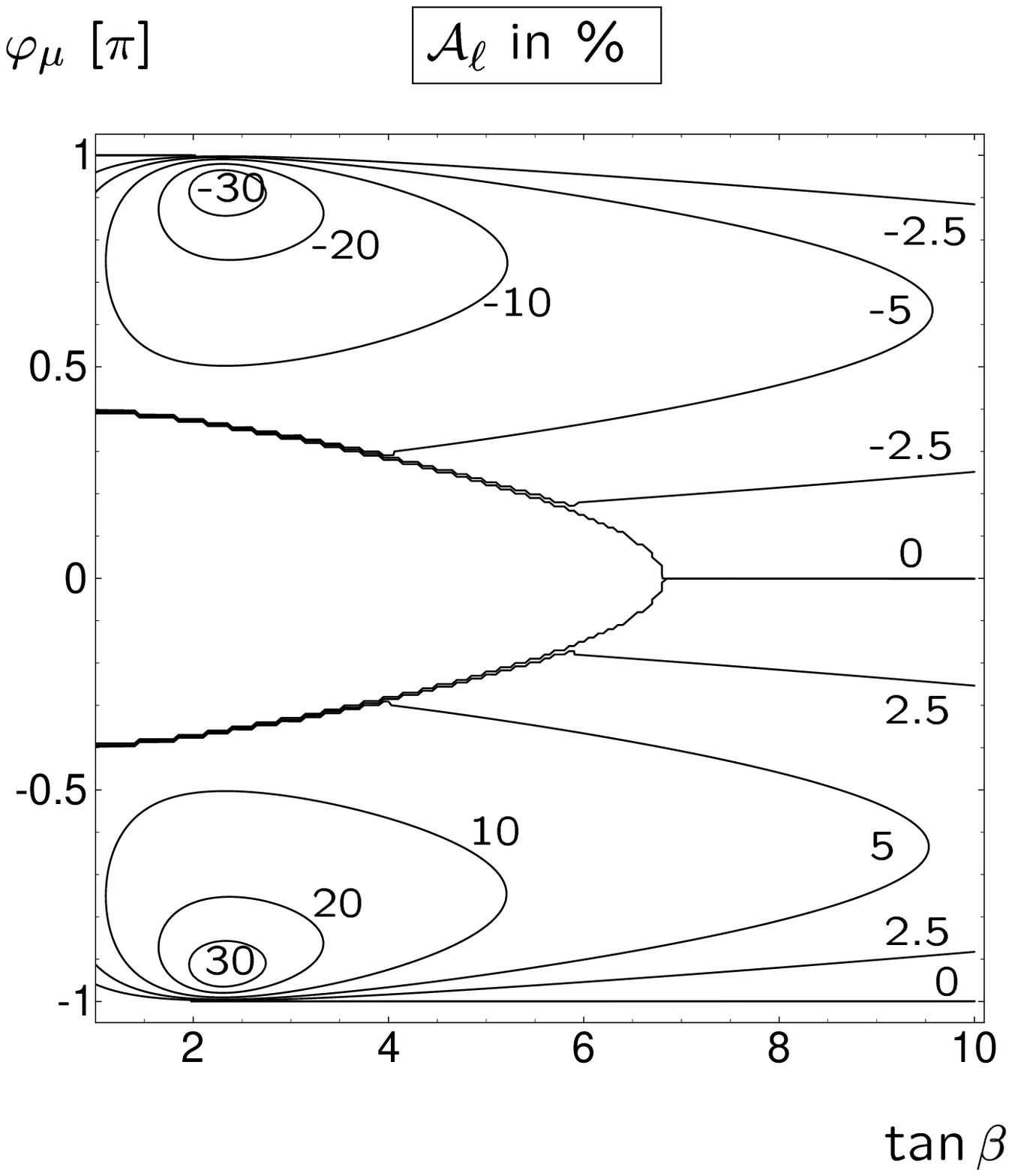,scale=0.3}}

\end{picture}
\caption{\label{fig:AT}
(a) T-odd asymmetry $A_T$
for $e^+ e^- \to \tilde{\chi}^0_1 \tilde{\chi}^0_2$,
$\tilde{\chi}^0_2 \to \tilde{\chi}^0_1 \ell^+ \ell^-$, $\ell = e,\mu$,
for $\sqrt{s}=500$ GeV (solid), $\sqrt{s}=350$ GeV (dashed) and
$P_{e^-}=-0.8$, $P_{e^+}=+0.6$
for $|M_1|=150$~GeV, $M_2=300$~GeV, $|\mu| = 200$~GeV,
$\phi_{\mu}=0$, $\tan\beta=10$, $m_{\tilde{\ell}_L}=267.6$~GeV,
$m_{\tilde{\ell}_R}=224.4$~GeV.
From \protect\cite{Bartl:2004jj}.
(b) Contours of the CP-odd asymmetry ${\cal A}_\ell$
for $e^+ e^- \to \tilde{\chi}^+_1 \tilde{\chi}^-_2$,
$\tilde{\chi}^+_1 \to \ell^+ \tilde{\nu}_\ell$, $\ell = e,\mu$,
for $\sqrt{s} = 800$~GeV, $P_{e^-} = -0.8$, $P_{e^+} = +0.6$ and 
$|\mu| = 400$~GeV, $M_2 = 200$~GeV, $m_{\tilde{\nu}} = 185$~GeV.
From \protect\cite{Bartl:2004vi}.
}
\end{figure}

In \cite{Bartl:2004jj} a T-odd asymmetry
\[
A_T = \frac{\sigma({\cal T}>0) - \sigma({\cal T}<0)}%
 {\sigma({\cal T}>0) + \sigma({\cal T}<0)}
 =
 \frac{\int {\rm sign}({\cal T}) |T|^2 d{\rm Lips}}%
 {{\int}|T|^2 d{\rm Lips}}
\]
is defined with help of the triple product
${\cal T}=\vec{p}_{\ell^+}\cdot (\vec{p}_{\ell^-}\times\vec{p}_{e^-})$
of the initial electron momentum $\vec{p}_{e^-}$ and
the two final lepton momenta $\vec{p}_{\ell^+}$ and $\vec{p}_{\ell^-}$,
where $T$ denotes the amplitude and ${\int}|T|^2 d{\rm Lips}$ is
proportional to the cross section $\sigma$ of the process
$e^+ e^- \to \tilde{\chi}^0_i \tilde{\chi}^0_j \to
\tilde{\chi}^0_i \tilde{\chi}^0_1 \ell^+ \ell^-$.
$A_T$ can be directly measured without
reconstruction of the momentum of the decaying neutralino or further
final-state analyses.
In representative scenarios of the SUSY parameters it has been shown
that the asymmetry can reach values $A_T = {\cal O}(10\,\%)$
(Fig.~\ref{fig:AT} (a)).

In \cite{Bartl:2004vi} a CP-odd asymmetry 
$A = \frac{1}{2}(A_T - \bar{A}_T)$, where $\bar{A}_T$ denotes the CP
conjugate of $A_T$,
in chargino production
$e^+ e^- \to \tilde{\chi}^+_i \tilde{\chi}^-_j$ with subsequent
two-body decay $\tilde{\chi}^+_i \to \ell^+ \tilde{\nu}_\ell$,
$\ell = e, \mu, \tau$,
is studied with help of the triple product
${\cal T}=\vec{p}_{\ell^+}\cdot
  (\vec{p}_{e^-}\times\vec{p}_{\tilde{\chi}^+_i})$.
To measure this asymmetry the momentum $\vec{p}_{\tilde{\chi}^+_i}$
of the decaying chargino has to be reconstructed.
$A$ is sensitive to $\phi_\mu$ and reaches values up to 30\,\% in some
regions of the parameter space (Fig.~\ref{fig:AT} (b)).

%***********************************************************************

\end{document}